# Voltage control of the quantum scattering time in InAs/GaSb/InAs trilayer quantum wells

M. Meyer[1,a)], S. Schmid[1], F. Jabeen[1], G. Bastard[1,2], F. Hartmann[1,b)] and S. Höfling[1]

[1]*Technische Physik, Physikalisches Institut and Würzburg-Dresden Cluster of Excellence ct.qmat, Am Hubland, D-97074 Würzburg, Germany.*

[2]*Physics Department, École Normale Supérieure, PSL 24 rue Lhomond, 75005 Paris, France*

We study the evolution of the quantum scattering time by gate voltage training in the topological insulator based on InAs/GaSb/InAs trilayer quantum wells. Depending on the minimal gate voltage applied during a gate voltage sweep cycle, the quantum scattering time can be improved by 50 % from 0.08 ps to 0.12 ps albeit the transport scattering time is rather constant around 1.0 ps. The ratio of the quantum scattering time versus transport scattering time scales linearly with the charge carrier density and varies from 10 to 30, indicating Coulombic scattering as the dominant scattering mechanism. Our findings may enable to improve the residual bulk conductivity issue and help in observing helical edge channels in topological insulators based on InAs/GaSb quantum well heterostructures even for macroscopic devices.

a) Email: manuel.meyer@physik.uni-wuerzburg.de
b) Email: fabian.hartmann@uni-wuerzburg.de

InAs/GaSb bilayer quantum wells (BQWs) are two dimensional topological insulators where the electron and hole gases are spatially separated in the InAs- and GaSb-layers, respectively[1]. Thu in contrast to the prototypal TI based on HgTe/CdTe quantum wells, a tuning between a trivial and topological phase by a dual gating approach in InAs/GaSb BQWs makes the material system interesting for potential device applications[2]. However, despite extensive studies on InAs/GaSb BQWs and indications for helical edge transport a fully convincing demonstration of these helical edge states is still elusive[3–9]. By adding another InAs-layer to the BQW, a symmetrical InAs/GaSb/InAs trilayer quantum well (TQW) can be formed[10–12]. TQWs enable significant higher band gap energies up to 60 meV for highly strained quantum wells enabling higher temperature operation with reduced bulk conductivity. However, even for large-gap topological insulators (TIs) in other material systems, e.g. $WTe_2$ ($E_{gap}$ = 100 meV), the conductance quantization could only be seen for channels of a few hundred nm[13]. Also, in the prototypal material system for TIs, HgTe/CdTe QWs helical edge transport could only be observed for channels of a few µm length[14]. This is mainly caused by scattering at small charged islands[15,16] formed by charged defects in the sample. Lunczer et al. however introduced a new measurement technique to overcome this bottleneck[17]. By sweeping the top-gate voltage to certain negative values and back, they showed that the gap conductance in the topological insulating gap could be increased for the optimal sweeping length. In addition, the scattering length increases to 175 µm. These improvements are assigned to a flattening of the potential landscape in the gap region by this gate-training method.

In this work, we study the evolution of the quantum scattering time $\tau_q$ in InAs/GaSb/InAs trilayer quantum wells by gate-voltage training. By sweeping the top gate from a maximum value $V_{max}$ to a minimum value $V_{min}$, we observe a hysteresis in the longitudinal resistance $R_{xx}$. Sweeps performed at low magnetic fields provide a distinct difference between the two hysteresis loops, where the amplitude of the Shubnikov-de Haas (SdH) oscillations increases. This is due to a change in the quantum scattering time for both sweeping directions. We investigate this with another experiment in which we change the gate sweeping length and observe a significant increase in the quantum scattering time whereas the transport scattering times remains constant. Furthermore, the ratio of $\tau_q$ and the transport scattering time $\tau_t$ allows us to determine the dominant scattering process.

For our experiments, we used an InAs/GaSb/InAs TQW (10/7/10 nm) with an indirect inverted gap of around $E_{gap}$ = 4 meV. Hall bars of constant width W = 20 µm but different lengths L = 20, 40, 60 and 80 µm (labeled S20, S40, S60 and S80, respectively) were processed. For further information on the growth and processing details see Ref. ([18]). All measurements were conducted at a temperature of 4.2 K in the dark. In Fig. 1(a) $R_{xx}$ as a function of the gate voltage $V_{TG}$ is shown for sample S60. During the gate voltage sweep, the magnetic field value was set constant to B = 3 T. The voltage sweep is performed by starting from the maximal voltage $V_{max}$ = +10 V down to the minimal gate voltage $V_{min}$ = -10 V (down-sweep, colored in red). Afterwards, the sweep direction changes and the gate voltage is increased back to $V_{max}$ (up-sweep, colored in black). For both sweep directions, a resistance peak ($V_{CNP,up}$ for the up-sweep and $V_{CNP,down}$ for the down-sweep) is visible when the Fermi energy is located in the gap region. For higher $V_{TG}$ the Fermi energy lies in the conduction band and for lower $V_{TG}$ it is in the valence band. Two prominent differences between the two sweep directions can be identified. First, the gate voltage of the gap region differs by a voltage difference ΔV between the up- and the down-sweep. This hysteresis occurs probably due to charge accumulation at the interface between the gate dielectric and the semiconductor[19,20]. Second, a prominent difference between the up- and down-sweep in the amplitude of the SdH-oscillations can be identified. Fig. 1(b) shows this in more detail. Here, the oscillations for the up- and down-sweep are depicted. In the graph, the voltage values are normalized to the charge neutrality point (CNP), $V_{TG} - V_{CNP}$, and the longitudinal resistance is displayed by subtracting the baseline resistance $R_{xx,base}$. The down-sweep oscillation (red) has amplitudes around $\Delta R_{xx,down} = 20 - 60\ \Omega$, whereas the values for the up-sweep (black) are $\Delta R_{xx,up} = 60 - 100\ \Omega$. The difference in the amplitudes at the same position is due to different quantum scattering times[21–23].

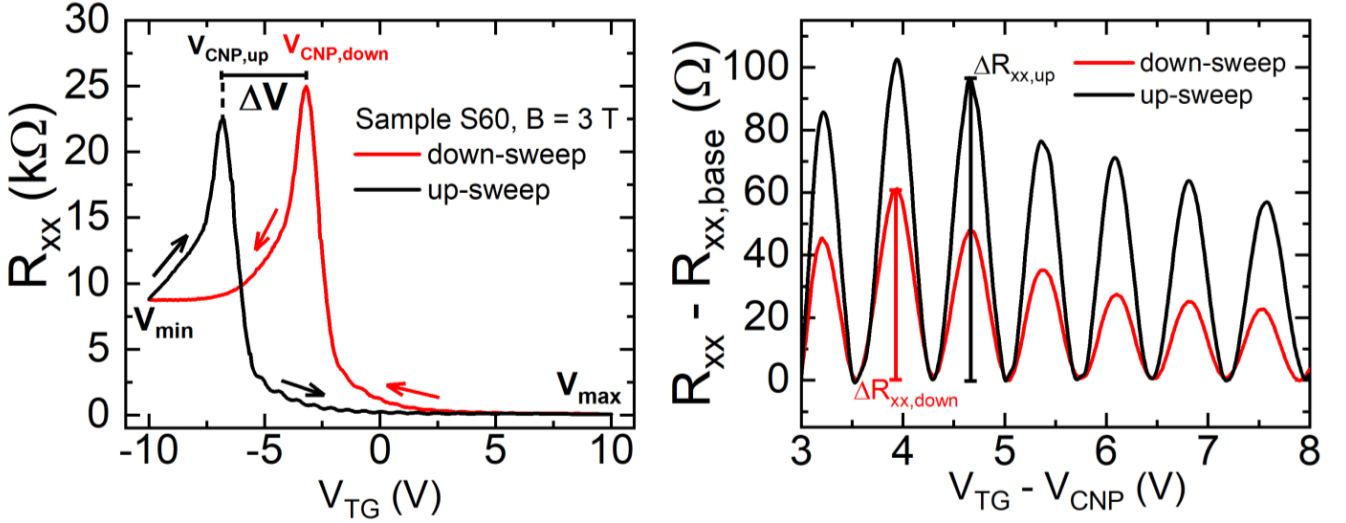

**Fig. 1. (a)** $R_{xx}$ as a function of the top-gate voltage $V_{TG}$ at $B = 3$ T for sample S60. The down- and up-sweep are color-coded in red and black, respectively. **(b)** SdH-oscillations for a normalized (to the CNP) top-gate voltage in the range of $V_{TG} - V_{CNP} = 3 - 8$ V. From the oscillations the baseline $R_{xx,base}$ is subtracted for comparison. Evidently, the observed amplitudes of the SdH oscillations for the up-sweep, $R_{xx,up}$, are more pronounced compared to the down-sweep, $R_{xx,down}$, direction.

In Fig. 2(a) the baseline corrected longitudinal resistance for S60 versus magnetic field strength is plotted for three different minimal gate voltages during a sweep cycle. For all traces, the gate voltage was swept from the maximal positive gate voltage of $V_{max} = +10$ V to the minimal gate voltages $V_{min} = -4.0$ V, $-4.5$ V, and $-5.0$ V. Afterwards the gate voltage was increased again and set constant at different $R_{xx}$-values for each $V_{min}$ corresponding to a fixed charge carrier density. Then the Hall traces were recorded. Please note that the gate voltage during the measurement may be slightly different but the charge carrier densities remain constant with $n \approx 6 \times 10^{11}\ cm^{-2}$. One observes that the SdH-oscillations appear at magnetic fields below 1.25 T for $V_{min}$ = -4.5 V and -4.0 V. In addition, the oscillation amplitude is largest for $V_{min}$ = -4.5 V. For $V_{min}$ = -5.0 V SdH-oscillations appear only at larger magnetic field values, i.e. well above 1.25 T and their amplitude is lower compared to the $V_{min}$ = -4.5 V and -4.0V values. From the amplitude of the SdH-oscillations, we extract the quantum scattering time for $V_{min}$ = 4.5 V as presented in Figures 2(b) and (c). Panel (b) shows an example of the baseline corrected $R_{xx}$ over 1/B for $V_{min}$ = -4.5 V with the envelope function plotted in orange dashed lines. From that, we extract the values $\Delta R_{xx}$ divided by a thermal factor $\gamma_{th}$[21] (black squares) for each amplitude as presented in panel (c). From the slope of the linear fit $\tau_q$ is extracted[21].

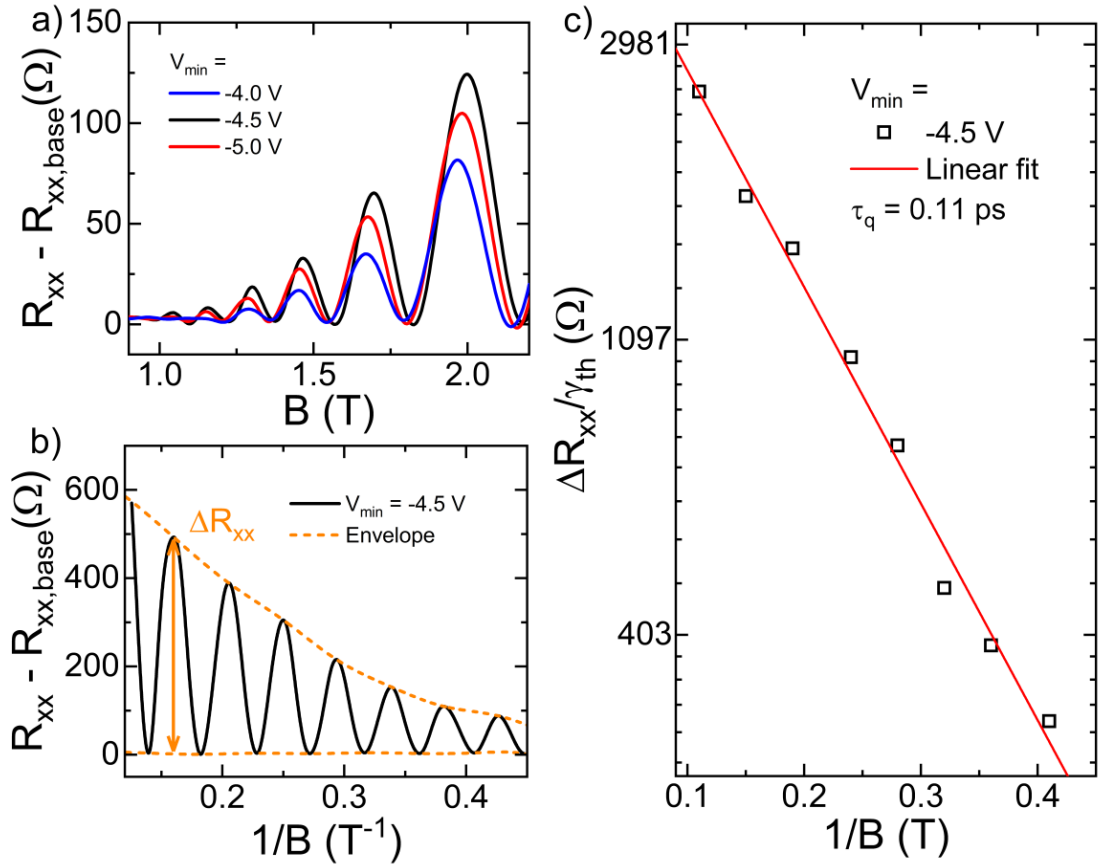

**Fig. 2.** (a) $R_{xx}$ (baseline corrected) in dependence of the magnetic field at $V_{min}$ = -4 V, -4.5 V and -5 V for sample S60. For $V_{min}$ = -4.5 V the oscillations appear at the lowest magnetic field values. For $V_{min}$ = -5.0 V, the onset of SdH-oscillations occur at the largest magnetic field value. (b) $R_{xx}$ (also baseline corrected) over 1/B for $V_{min}$ = -4.5

V. The envelope function is plotted in orange dashed lines. This is used to extract the values $\Delta R_{xx}$ for each amplitude. In **(c)** the $\Delta R_{xx}$ values divided by a thermal factor $\gamma_{th}$ (black squares) are shown. From the slope of the linear fit $\tau_q$ is extracted.

The measurement procedure as explained in Fig. 2 was then carried out for all four samples with changing minimal gate voltages $V_{min}$ = -3 to -10 V in steps of 1V (for S60 two additional measurements for $V_{min}$ = -4.5 V and -5.5 V were added) to extract $\tau_q$. For clarification, more details on the measurement procedure can be found in the Supplemental Material[24]. Furthermore, the transport scattering time $\tau_t$ was also extracted for comparison. The measurement range was used that the Fermi energy passes through the CNP and hence the gate training via $V_{min}$ can affect the potential in the gap region and the operation condition where topological edge states appear[17]. For $V_{min} \geq$ -3 V we do not see a resistance peak in the up-sweep for most samples. Every measurement was performed at a fixed starting point regarding the charge carrier density with n ≈ $6\times10^{11}$ cm$^{-2}$. This allows us to compare each measurement from each sample. All four samples show a similar trend for $\tau_q$ in dependence on $V_{min}$: Starting at $V_{min}$ = -3 V, $\tau_q$ is increasing with decreasing $V_{min}$. For all samples, $\tau_q$ peaks between $V_{min}$ = -4 and -5 V with $\tau_q$ = 0.11 – 0.12 ps. For clarity, we added the initial $\tau_q$ (when sweeping directly from $V_{max}$ to the starting position without gate training) as dashed lines for each sample and a detailed comparison with and without gate training can be found in Fig. S4 in the Supplemental Material[24]. Further reducing $V_{min}$ leads to decreasing $\tau_q$ until a minimal value can be seen for each sample at a certain $V_{min}$. As provided in the Supplemental Material[24], the decrease of $\tau_q$ below $V_{min}$ < -5 V correlates with the appearance of the hysteresis. The hysteresis is typically assigned to charge trapping at the semiconductor/oxide interface[17,19]. For $V_{min} \geq$ -5 V and when $\tau_q$ is increasing, no hysteresis is observable (see also Fig. S2 from the Supplemental Material[24]). Therefore, we assume that the improvement of $\tau_q$ is not related to defect states at the surface but related to defects in the surrounding barrier material. As we show later, we can estimate the defect density and their spacing to the two dimensional electron gas. We assume that the improvement of $\tau_q$ is based on the reduction of charged defects that are close to the Fermi energy in the surrounding barrier material[25,26]. While we can observe a prominent improvement of the quantum scattering time, the transport scattering time remains rather constant by variations of $V_{min}$ (squares in Fig. 3). Therefore, $\tau_t$ is nearly unaffected by the gate training excluding thus a simple screening of impurities. For all samples we observe a significant improvement of $\tau_q$ up to 50%. The change of $\tau_q$ shows that indeed via gate voltage training the quantum level broadening $\Gamma=\hbar/2\tau_q$ can be reduced, which is consistent with a flattening of the potential landscape[17]. Therefore, the gate training is useful to improve the transport properties of our heterostructure.

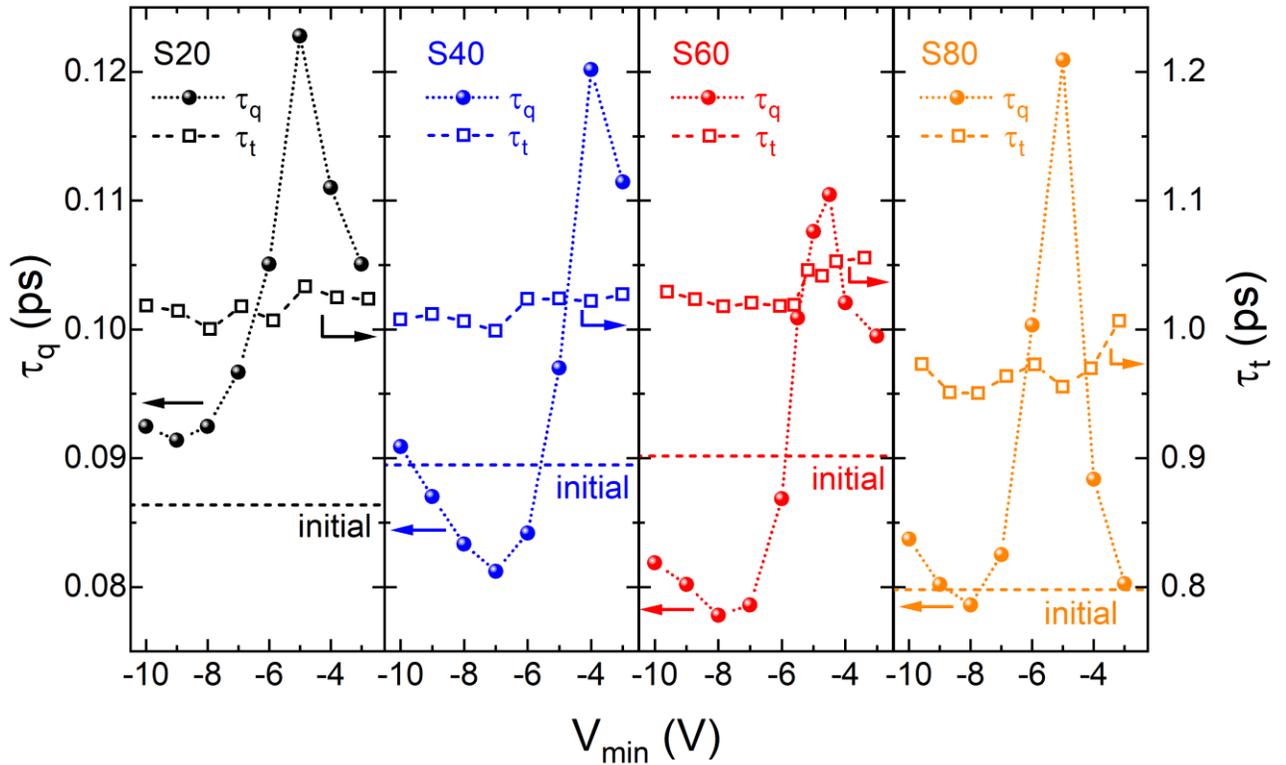

**Fig. 3.** Quantum and transport scattering time as a function of the gate sweeping length $V_{min}$ = -3 to -10 V for the four different samples. All samples show a similar trend: Starting from $V_{min}$ = -3 V the quantum scattering time increases with larger gate sweeping lengths. A maximum is observable around $V_{min}$ = -4 to -5 V with $\tau_{q,max} \approx 0.12$

ps (for S60 $\tau_{q,max} \approx 0.11$ ps). Afterwards, $\tau_q$ decreases until a minimum value for each sample and then starts to increase again. $\tau_t$ is nearly constant and therefore unaffected by the effects of the gate training. The initial $\tau_q$ (when sweeping from $V_{max}$ directly to the starting position without gate training) is also added for each sample as a dashed line.

Besides the dependency on $V_{min}$, $\tau_q$ is also dependent on the charge carrier density[27]. Figure 4 displays the transport scattering time $\tau_t$ (in (a)), the quantum scattering time $\tau_q$ (in (b)), and the ratio of both time scales, $\tau_t/\tau_q$ (in (c)), as a function of the charge carrier density from $4\times10^{11}$cm$^{-2}$ to $14\times10^{11}$cm$^{-2}$. The measurement was performed similarly to the previous ones with the gate voltage sweep starting at $V_{max}$ = 10 V, sweeping it to a fixed minimal gate voltage $V_{min}$ = -10 V, and then back to the measurement voltage $V_{xx,start}$. This voltage $V_{xx,start}$ was changed between the measurements, which corresponds to different charge carrier densities.

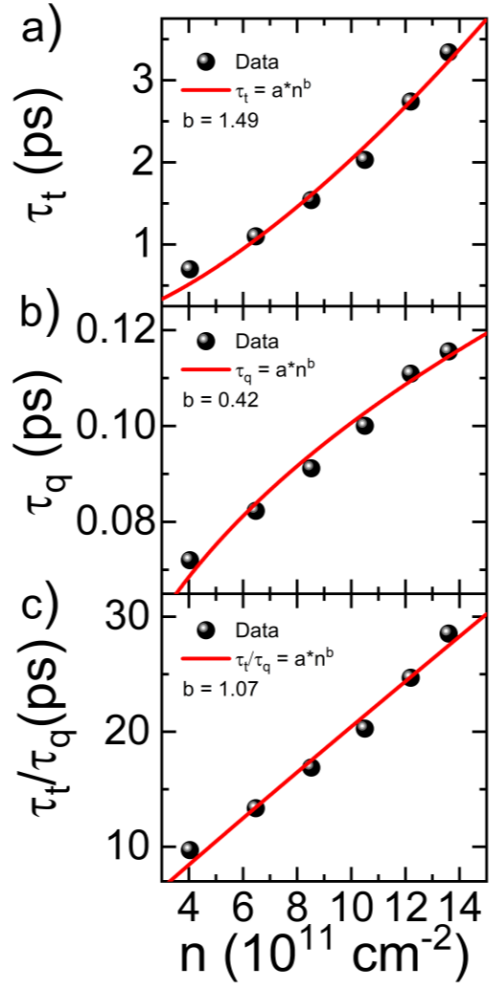

**Fig. 4. (a)** Transport scattering time $\tau_t$, **(b)** quantum scattering time $\tau_q$ and **(c)** the ratio $\tau_t/\tau_q$ as a function of the charge carrier density. $\tau_t$, $\tau_q$ and $\tau_t/\tau_q$ were fitted with a power law with b = 1.49, 0.42 and 1.07, respectively.

The transport scattering time $\tau_t$ increases from 0.7 ps up to 3.5 ps and the quantum scattering time $\tau_q$ increases from 0.07 ps up to 0.12 ps. A power law dependence $n^b$ for both scattering times can be seen and the fitting enables us to extract powers of b = 1.49 and 0.42, for the transport and quantum scattering time, respectively. Therefore, the experimental ratio $\tau_t/\tau_q$ increases almost linearly with b = 1.07. In our trilayer well system, the positions (in either InAs- or GaSb-layers or in AlAsSb-barriers) and natures (acceptors, donors) of the dopants are unknown. Therefore, it is impossible to make an accurate modeling of the quantum and transport scattering times versus the 2D charge concentration n. The experiments (see

Fig. 4) show that the latter is much longer than the former, which eliminates the possibility that the dominant static scatterers are short range. Most likely, they are charged impurities or traps. In addition, plotting their ratio against n shows an almost linear behavior over a large range of n values. We build a simplified model giving explicit predictions of the n variations of the relaxation times and their ratio[28]. The assumptions are that all the ionized impurities stay on a plane located at $|z_0|$ far from the planar 2D gas located on the plane z = 0. The screening is treated at the Thomas Fermi approximation and T = 0 K. Then, we get for $\tau_q$ and $\tau_t$:

$$\frac{\hbar}{\tau_q} = \frac{m_e N_d}{\pi \hbar^2} \left(\frac{2\pi e^2}{\kappa}\right)^2 \int_0^\pi d\theta \frac{1}{(q_0 + 2k_F \sin\frac{\theta}{2})} \exp(-4k_F |z_0| \sin\frac{\theta}{2}) \quad (1a)$$

$$\frac{\hbar}{\tau_t} = \frac{m_e N_d}{\pi \hbar^2} \left(\frac{2\pi e^2}{\kappa}\right)^2 \int_0^\pi d\theta \frac{(1-\cos\theta)}{(q_0 + 2k_F \sin\frac{\theta}{2})} \exp(-4k_F |z_0| \sin\frac{\theta}{2}) \quad (1b)$$

where $\kappa = 4\pi\varepsilon_0\varepsilon_r$ with $\varepsilon_0$, $\varepsilon_r$ being the vacuum and relative dielectric constants, respectively. $k_F$ is the Fermi wavevector, $q_0 = 2/a_B$ where $a_B$ is the effective Bohr radius and $N_d$ the 2D impurity concentration. $\theta$ is the angle between the incident and scattered wavevectors. The two angular integrals $I_q$ and $I_t$ can be transformed to get:

$$I_q = \frac{1}{2k_F |z_0| q_0^2} \int_0^{4k_F |z_0|} dt \frac{e^{-t}}{(1+\frac{t}{2|z_0|q_0})^2 \sqrt{1-(\frac{t}{4k_F |z_0|})^2}} \quad (2a)$$

$$I_t = \frac{1}{2k_F |z_0| q_0^2} \frac{1}{8k_F^2 z_0^2} \int_0^{4k_F |z_0|} dt \frac{t^2 e^{-t}}{(1+\frac{t}{2|z_0|q_0})^2 \sqrt{1-(\frac{t}{4k_F |z_0|})^2}} \quad (2b)$$

Now the model is completed by assuming both $k_F|z_0|$ and $q_0|z_0|$ are very large. Then $I_q \approx \frac{1}{2k_F|z_0|q_0^2}$ and $I_t \approx \frac{1}{2k_F|z_0|q_0^2} \frac{1}{4k_F^2 z_0^2}$, thus $\frac{\tau_t}{\tau_q} \approx 4k_F^2 z_0^2$. Therefore, the model predicts that $\tau_q$ scales as $n^{1/2}$ and $\tau_t$ scales as $n^{3/2}$. Their ratio is thus linearly dependent on n. These predictions qualitatively agree with our experiments. From the model, we can also extract the defect densities $N_d \approx 0.4 \times 10^{11}$ cm$^{-2}$ and the average scattering

plane distance of these defect states $z_0 \approx 9$ nm. Again, this points towards charged defects in the surrounding barrier material close to our TQW as the most prominent origin of scattering.

In summary, we evaluated the evolution of the quantum scattering time $\tau_q$ in InAs/GaSb/InAs TQWs by gate voltage training. By sweeping the top-gate voltage from a maximum value $V_{max}$ to a minimum value $V_{min}$, we observe a hysteresis in the longitudinal resistance. Sweeps performed at low magnetic fields provide a distinct difference between the two hysteresis loops, where the amplitude of the Shubnikov-de Haas oscillations increases from the down-sweep to the up-sweep. This is due to a change in the quantum scattering time for both sweeping directions. We investigate this with another experiment in which we change the gate sweeping length and observe an increase in the quantum scattering time of about 50% that corresponds to a decrease of the quantum level broadening whereas the transport scattering time is rather unaffected. We emphasize that this will help in observing helical edge channels in topological insulators based on InAs/GaSb heterostructures even for macroscopic devices. Furthermore, the ratio of $\tau_q$ and the transport scattering time $\tau_t$ shows that long range Coulombic scattering is the dominant scattering mechanism and that defects in the barriers close to the TQW are the most prominent origin of scattering.

The work was supported by the Elite Network of Bavaria within the graduate program "Topological Insulators". Expert technical assistance by A. Wolf and M. Emmerling is gratefully acknowledged.

# Supplemental Material for 'Voltage control of the quantum scattering time in InAs/GaSb/InAs trilayer quantum wells'

M. Meyer[1], S. Schmid[1], F. Jabeen[1], G. Bastard[1,2], F. Hartmann[1] and S. Höfling[1]

[1]*Technische Physik, Physikalisches Institut and Würzburg-Dresden Cluster of Excellence ct.qmat, Am Hubland, D-97074 Würzburg, Germany.*

[2] *Physics Department, École Normale Supérieure, PSL 24 rue Lhomond, 75005 Paris, France*

**Measurement procedure**

Fig. S1 shows the measurement procedure details. Starting from $V_{max}$ (always +10 V), $V_{TG}$ is decreased to $V_{min}$ and directly increased again to a set starting value $R_{xx,start}$. $R_{xx,start}$ is the same for every $V_{min}$ (so we have the same charge carrier density for every Hall measurement) whereas the gate-voltage value $V_{TG,start}$ changes for every $V_{min}$. As an example: For $V_{min}$ = -9 V, $R_{xx,start}$ = 0.48 kΩ and $V_{TG,start}$ = -2.95 V. For $V_{min}$ = -8 V, $R_{xx,start}$ = 0.48 kΩ (the same as for $V_{min}$ = -9 V) but $V_{TG,start}$ = -1.94 V. At $V_{TG,start}$, we then perform the Hall measurement and extract the quantum scattering time $\tau_q$ (and the transport scattering time $\tau_t$) out of the SdH-oscillations. Therefore, we get $\tau_q$ versus the gate sweeping length $V_{min}$. Additionally, between each measurement a whole gate-voltage sweep from $V_{TG}$ = +10 V to -10 V and back is performed to erase the effects from the previous measurements and to ensure the same initial conditions for each measurement.

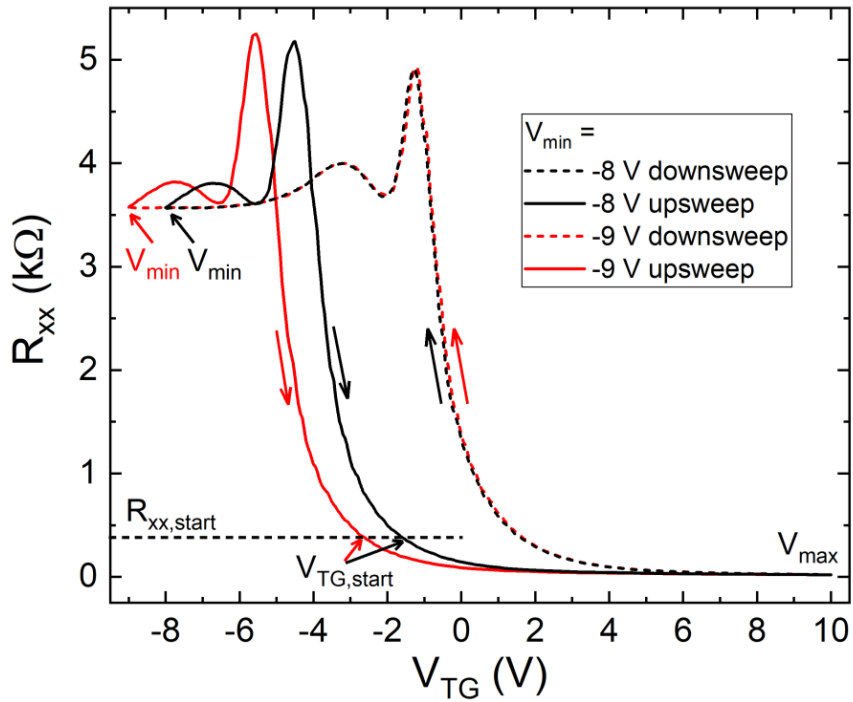

**Fig. S1.** Scheme for the detailed explanation on the measurement procedure. Dependent on $V_{min}$, the gate voltage used for the Hall measurement, $V_{TG,start}$, differs as we perform the experiment at fixed initial $R_{xx,start}$ values.

**Connection between $\tau_q$ and the hysteresis**

The dependence of $\tau_q$ on $V_{min}$ is connected to the hysteresis. Therefore, we measured the peak position for the up- and down-sweep ($V_{peak,up}$ and $V_{peak,down}$) versus $V_{min}$ (Fig. S2 (a)) and compared the hysteresis width $\Delta V = V_{peak,down} - V_{peak,up}$ to $\tau_q$ (Fig. S2 (b)) for S40. When no hysteresis is present (for this case $V_{min} \geq -4$ V but this changes slightly for each sample), $\tau_q$ is increasing. Afterwards, the hysteresis increases and $\tau_q$ decreases. Due to the hysteresis being caused by accumulated charges (probably in the oxide or semiconductor/oxide interface[1]) we assume the decrease in $\tau_q$ is caused by these accumulated charges which are also the reason for the hysteresis.

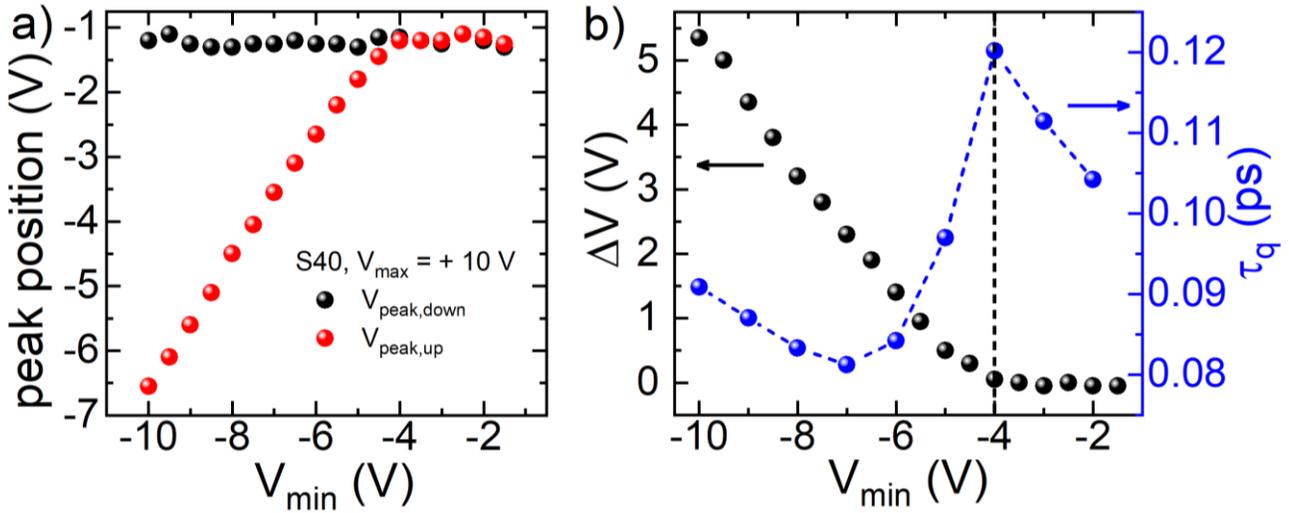

**Fig. S2. (a)** Evolution of the peak positions of the up- and down-sweep ($V_{peak,up}$ and $V_{peak,down}$) for a variation of $V_{min}$ and a fixed $V_{max} = +10$ V for S40. **(b)** Hysteresis width $\Delta V$ and $\tau_q$ versus $V_{min}$. When $\tau_q$ is increasing the hysteresis remains nearly constant. Afterwards, $\tau_q$ decreases and a pronounced increase of $\Delta V$ is observed.

**Quantum scattering time with and without gate training**

The usefulness of the gate training can be further proved by comparing the quantum scattering time with and without gate training as shown in Fig. S3 for sample S40. Without gate training, we choose two different configurations:

1. We cool down the sample, sweep directly to the starting voltage point (where n ≈ 6 × 10 cm$^{-2}$) and then perform the Hall measurement to extract $\tau_q$ (face-up triangles).

2. We cool down the sample, sweep to $V_{max}$ = +10 V, sweep back to the starting voltage and then perform the Hall measurement to extract $\tau_q$ (face-down triangles).

For both configurations, we observe comparable quantum scattering times with $\tau_q$ ≈ 0.091 ps (for the first configuration) and $\tau_q$ ≈ 0.096 ps (for the second configuration). With gate training on the other hand, we observe quantum scattering times of up to $\tau_q$ ≈ 0.129 ps (circles). Thus, the gate training improves the initial quantum scattering time by about 30 %.

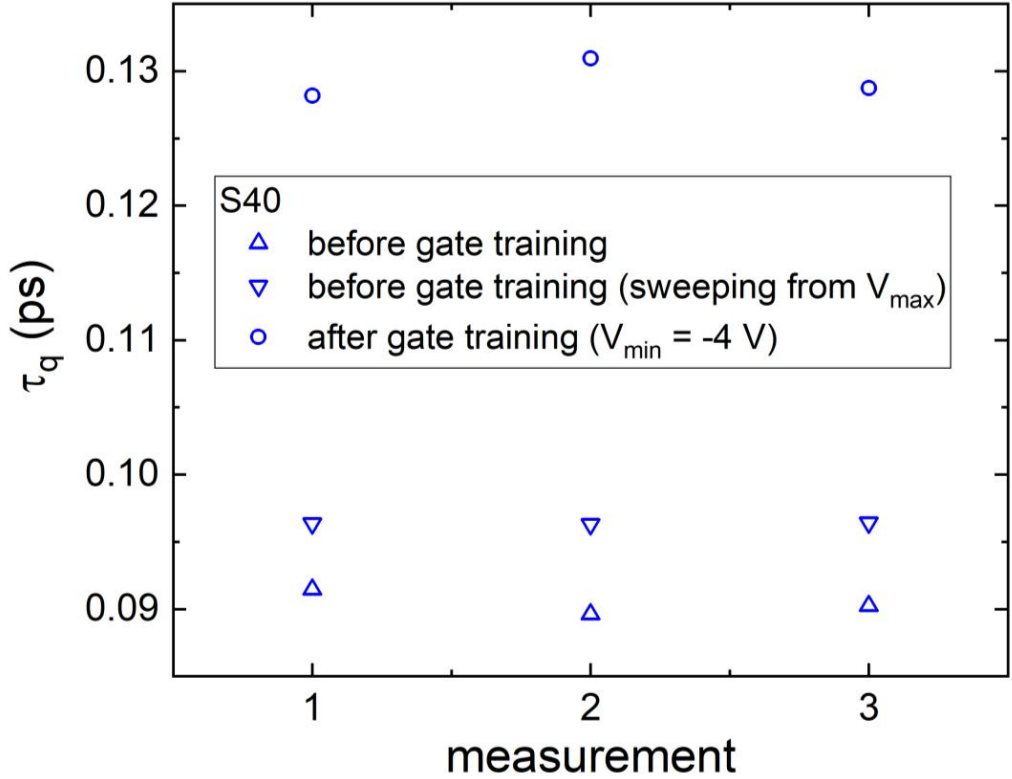

**Fig. S3.** $\tau_q$ before gate training (face-up triangles), before gate training but sweeping first to $V_{max}$ (face-down triangles), and after gate training (circles) for the same starting point corresponding to n ≈ 6 × 10$^{11}$ cm$^{-2}$ for S40. The measurements were performed three times, i.e. cooling down the sample. A clear improvement with gate training in comparison to the two configurations without gate training is observable.

**Time dependence of the quantum scattering time**

Due to the dependence of $\tau_q$ on $V_{min}$ being linked to the hysteresis and therefore some kind of charge transfer we performed measurements regarding the time dependence at S40. This was done to see how stable our system is and how much $\tau_q$ changes with time at a certain $V_{min}$. Fig. S4 (a) depicts the time dependent results for $R_{xx}$ for different $V_{min}$. For this measurement, we sweep $V_{TG}$ from $V_{max}$ = +10 V to the different $V_{min}$ = -3 V to -10 V and measure $R_{xx}$ for t = 66 s. Between each measurement we additionally performed a whole sweep from $V_{TG} = +10$ V to -10 V and back. The maximum relative change in $R_{xx}$ ($R_{xx,start} = R_{xx}(t = 0$ s)) is about 4%. For further evaluation of the time dependence we used $V_{min} = -3$ V and measured $\tau_q$ versus the waiting time at this point. Starting at $V_{max} = +10$ V we swept to $V_{min}$ and waited for $t = 0$ to 480s. Afterwards, $V_{TG}$ was set to a starting value $V_{xx,start}$ (similar to the measurement in the manuscript where $n \approx 6 \times 10^{11}$ cm$^{-2}$) and the Hall measurements to calculate $\tau_q$ were performed. The results are shown in Fig. S4 (b).

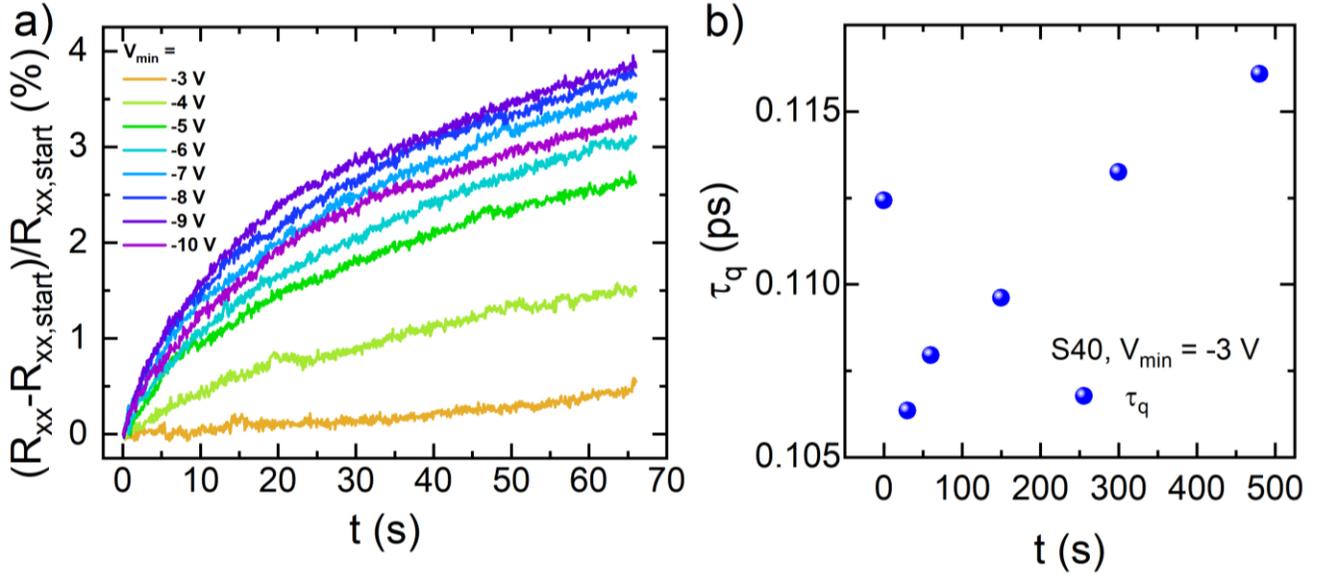

**Fig. S4.** (a) Relative change of $R_{xx}$ ($R_{xx,start} = R_{xx}(t=0s)$) versus time for different $V_{min}$ from -3 V to -10 V for S40. (b) $\tau_q$ for $V_{min} = -3$ V versus the waiting time at this position for $t = 0$ s to 480 s. At first, $\tau_q$ is decreasing but afterwards it is increasing for longer waiting times.

$\tau_q$ decreases until $t = 30$ s with a value of $\tau_{q,min} = 0.106$ ps. Afterwards it is increasing again until its maximum value of $\tau_q = 0.116$ ps at $t = 480$ s. The relative change here is about 5%. Although there is some time dependency, the overall changes especially for $\tau_q$ are a lot smaller compared to the changes achieved by gate training (up to 50%).